%% file: _main.tex
\documentclass{article}

\usepackage{_style}

\usepackage[utf8]{inputenc} 
\usepackage[T1]{fontenc}    
\usepackage{microtype}      
\usepackage{authblk}        

\usepackage{hyperref}       
\usepackage{booktabs}       
\usepackage{multicol}
\usepackage{multirow}
\usepackage{makecell}
\usepackage{amsfonts}       
\usepackage{amsmath}
\usepackage{graphicx}       
\usepackage{caption}
\usepackage{subcaption}
\usepackage[switch]{lineno} 
\usepackage[numbers,sort&compress]{natbib}
\usepackage{cleveref}
\usepackage{xcolor}
\usepackage{float}

\title{HistoEncoder: a digital pathology foundation model for prostate cancer}

\input{authors}

\begin{document}

\twocolumn[\begin{@twocolumnfalse}\maketitle\end{@twocolumnfalse}]

{
  \renewcommand{\thefootnote}{* }
  \footnotetext[1]{
        These authors contributed equally.\\
        \hspace*{1.15em}$\dagger$ Correspondence:\\
        \hspace*{1.9em}\href{mailto:esa.pitkanen@helsinki.fi}{esa.pitkanen@helsinki.fi}, \href{mailto:tuomas.mirtti@helsinki.fi}{tuomas.mirtti@helsinki.fi}\\
        \hspace*{1.8em}Code: \url{https://github.com/jopo666/HistoEncoder}
    }
}

\begin{abstract}
    \textbf{\normalsize\input{abstract.tex}}
\end{abstract}

\renewcommand\labelitemi{\tiny$\bullet$}

\input{manuscript.tex}

\input{extras.tex}

\bibliographystyle{unsrt}  
\bibliography{references}

\setcounter{table}{0}
\setcounter{figure}{0}
\renewcommand{\thetable}{S\arabic{table}}
\renewcommand{\thefigure}{S\arabic{figure}}
\clearpage
\onecolumn

\input{supplementary}

\end{document}

%% file: authors.tex
\author[1,*]{Joona Pohjonen}
\author[1,2,*]{Abderrahim-Oussama Batouche}
\author[1,3,4]{Antti Rannikko}
\author[1,5]{Kevin Sandeman}
\author[1]{Andrew Erickson}
\author[7,6,4,$\dagger$]{Esa Pitkänen}
\author[1,4,8,$\dagger$]{Tuomas Mirtti}

\affil[1]{Research Program in Systems Oncology, Faculty of Medicine, University of Helsinki}
\affil[2]{Doctoral Program in Computer Science, University of Helsinki}
\affil[3]{Department of Urology, Helsinki University Hospital }
\affil[4]{iCAN Digital Precision Cancer Medicine Flagship, Finland}
\affil[5]{Department of Pathology, Division of Laboratory Medicine, Skåne University Hospital, Malmö, Sweden}
\affil[6]{Research Program in Applied Tumor Genomics, Faculty of Medicine, University of Helsinki}
\affil[7]{Institute for Molecular Medicine Finland (FIMM), HiLIFE, University of Helsinki}
\affil[8]{Department of Pathology, Helsinki University Hospital }

%% file: abstract.tex
Foundation models are trained on massive amounts of data to distinguish complex patterns and can be adapted to a wide range of downstream tasks with minimal computational resources.
Here, we develop a foundation model for prostate cancer digital pathology called HistoEncoder by pre-training on 48 million prostate tissue tile images.
We demonstrate that HistoEncoder features extracted from tile images with similar histological patterns map closely together in the feature space.
HistoEncoder outperforms models pre-trained with natural images, even without fine-tuning or with 1000 times less training data. 
We describe two use cases that leverage the capabilities of HistoEncoder by fine-tuning the model with a limited amount of data and computational resources. 
First, we show how HistoEncoder can be used to automatically annotate large-scale datasets with high accuracy.
Second, we combine histomics with commonly used clinical nomograms, significantly improving prostate cancer-specific death survival models.
Foundation models such as HistoEncoder can allow organizations with limited resources to build effective clinical software tools without needing extensive datasets or significant amounts of computing.

%% file: manuscript.tex
\input{sections/1-intro}

\input{sections/2-methods}

\input{sections/3-results}

\input{sections/4-discussion}

%% file: sections/1-intro.tex
\section{Introduction}
\label{sec:intro}

Neural network-based solutions~\cite{deep_learning} have achieved impressive results with tissue diagnostics, often surpassing human counterparts in consistency, speed and accuracy~\cite{path_to_clinic, new_generation, ai_and_pathology}. 
For instance, a multiple instance learning model by Esteva {\em et al.}, trained and validated with clinical trial histological images and data, showed that prostate morphological features learnt by the model contain predictive information beyond conventional nomograms~\cite{Esteva2022}. A subsequent AI biomarker, predicting the benefit of adding androgen deprivation therapy to radiation therapy, primarily includes conventional Gleason patterns and image features extracted with a neural network as key components of the model~\cite{Spratt2023}.  

Despite promising results, recent work has demonstrated that neural networks perform substantially worse on datasets not used during training~\cite{nagpal, stain_norm, metastasis, campanella, melanoma}. A major driver of this performance reduction is dataset shift~\cite{dataset_shift, strongaugment}, where neural networks fail to generalize to data from a new clinical setting that differs from the training data. Fine-tuning pre-trained neural networks for downstream tasks has been shown to improve model robustness and reduce uncertainty~\cite{pre_training}. Many recent publications have confirmed this by leveraging neural networks pre-trained on natural images~\cite{natural_images_1, natural_images_2,natural_images_3,natural_images_4,natural_images_5,natural_images_6,natural_images_7,natural_images_8,natural_images_9,natural_images_10,natural_images_11, natural_images_12}. Still, due to the considerable domain shift between histological and natural images, transfer learning from a neural network pre-trained with natural images offers little benefit to performance \cite{transfer_learning_from_natural_images}. Thus, there is a need for neural networks pre-trained with histological images.

Recent advances in self-supervised learning \cite{mae, simmim,ibot, dino, dinov2} have resulted in the emergence of foundation models \cite{foundation_models, caron_foundation, goyal_foundation_1, goyal_foundation_2}. Foundation models are trained on large-scale datasets, leveraging unlabelled samples via self-supervised learning \cite{ssl_in_pathology1, ssl_in_pathology2}, and can be then easily adapted for downstream tasks, even without any additional training \cite{dino, dinov2}. 

\begin{figure*}
    \centering
    \includegraphics[width=0.60\textwidth]{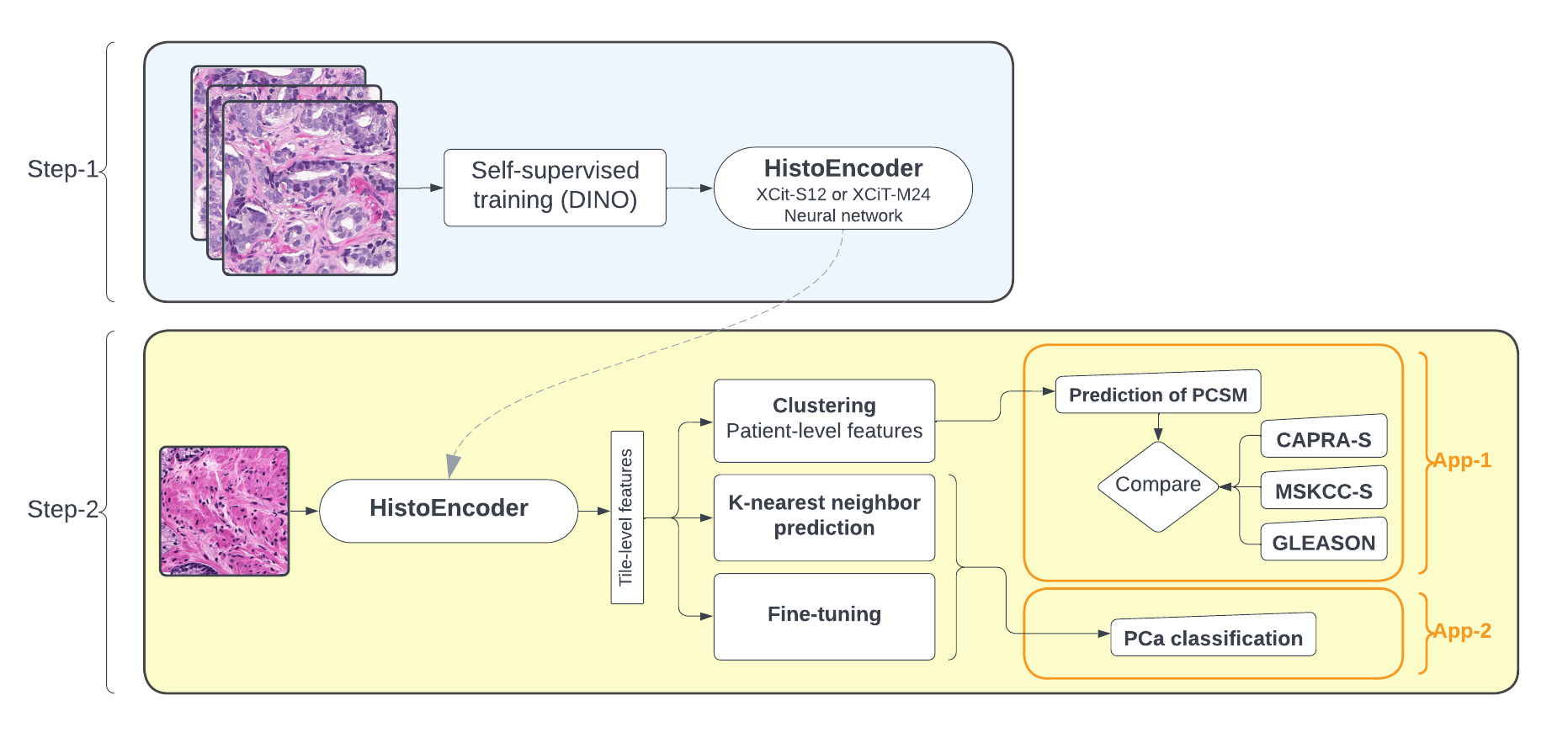}
    \caption{An overview of the HistoEncoder workflow. HistoEncoder utilizes the cross-covariance transformer (XCiT) as the backbone network. The models are pre-trained with 48 million prostate tissue tile images from 1,307 patients in a self-supervised manner (Step 1), and fine-tuned to cancer classification and mortality prediction tasks, or used without fine-tuning via a KNN classifier (Step 2).}
    \label{fig:overview}
\end{figure*}

Here, we train foundation models on 48 million tile images from thousands of histological slide images with prostate tissue (\cref{fig:overview}). These foundation models outperform models pre-trained with natural images by a large margin. Additionally, we describe two workflows leveraging foundation models. The first workflow describes a method for automatically annotating large-scale tissue image datasets, which we evaluate by annotating the largest publicly available histological dataset \cite{panda} with high accuracy. The second workflow integrates annotated significant histology features with commonly used clinical prognostic nomograms to improve the prediction of prostate cancer-specific mortality.

%% file: sections/2-methods.tex
\section{Materials and methods}
\label{sec:methods}

\input{tables/datasets}

\subsection{HistoEncoder models}
\label{sec:encoder_models}

In this study, we train cross-covariance image transformers (XCiT) \cite{xcit} with a self-supervised learning method DINO \cite{dino, dinov2}, which leverages discriminative signals between groups of images to learn useful features using self-distillation. Instead of focusing on maximizing the fine-tuning performance on a downstream task \cite{mae, simmim}, the encoder model training aims to maximize the representativeness of extracted features \cite{dinov2}. We selected XCiT due to linear complexity in the number of tokens,  allowing efficient processing of high-resolution images \cite{xcit}, which are commonly encountered in digital pathology. Two XCiT backbones are trained on prostate tissue samples, \texttt{prostate-s} and \texttt{prostate-m}, based on the small and medium-sized XCiT model variants  \texttt{XCiT-S12} and \texttt{XCiT-M24}, respectively.

Many recent models for clinical digital pathology image analysis have been pre-trained with natural images, often using supervised training methods~\cite{natural_images_1,natural_images_2,natural_images_3,natural_images_4,natural_images_5,natural_images_6, natural_images_7,natural_images_8,natural_images_9,natural_images_10,natural_images_11, natural_images_12}. Although it is common to use models pre-trained with supervised training methods, this does not always produce useful embedded features, when compared to models trained with self-supervised methods (see \cite{dino} and Supplementary Tab. 2, 3, and 4)
. Thus, to make comparisons between models trained with histological images and natural images more fair, all comparisons are against XCiT-S12 and XCiT-M24 models trained on natural images using the self-supervised method DINO \cite{dino, imagenet}. These models are denoted as \texttt{natural-s} and \texttt{natural-m}, respectively.

\subsection{Workflows leveraging encoder models}
\label{sec:pipelines}

We describe two workflows that leverage the image features from our encoder models. The first is for automatic pre-annotation of large imaging datasets, and the second is for combining histomics data with other data modalities. Command line interfaces and a Python module for running these workflows can be found from \url{https://github.com/jopo666/HistoEncoder}.


\subsubsection{Automatic annotation of large-scale slide image datasets}
\label{sec:automatic_annotation}

First, all slide images in the dataset are cut into small tile images, for example with \texttt{HistoPrep} \cite{histoprep}. Second, features for all tile images are extracted with either \texttt{prostate-s} or \texttt{prostate-m} encoder models. Third, all extracted features for the tile images are clustered into $n$ clusters with mini-batch K-means. Given that HistoEncoder produces similar embedded features for tile images with similar histological patterns, clusters can be labelled based on visual inspection of the prevailing histological pattern.
After automatically pre-annotating all tile images in a dataset with HistoEncoder, a pathologist can visually inspect a subset of the tile images in each cluster. Now, if a given cluster contains only a single histological pattern, such as stroma or epithelium, all tile images assigned to this cluster can be labelled based on the prevailing pattern. If a cluster contains too much variability, the tile images in this cluster can be further clustered and labelled to produce clusters with less variability. The process can then be repeated until enough tile images have been labelled for the task at hand.

\subsubsection{Combining histomics with other data modalities}
\label{sec:multimodality}

Here, we present a workflow to integrate HistoEncoder image features with other data modalities.
When additional data modalities, like spatial transcriptomics data \cite{spatial_transcriptomics} or tissue microarray (TMA) spot labels, provide information for smaller tissue regions, feature vectors can be directly generated for these regions using the \texttt{prostate-s} or \texttt{prostate-m} encoder models. 

In contrast, if the data modalities to be integrated contain information for larger tissue regions or whole or multiple slide images, the slide images must be cut into smaller tile images that are compatible with HistoEncoder. A common example is patient-level clinical data, where data from each patient pertains to all tile images from the same patient. 
In this case, we will first automatically annotate and cluster all tile images following the workflow in \cref{sec:automatic_annotation}. Each cluster contains tile images with similar histological patterns. Second, we will calculate, for each patient, the proportion of tile images assigned to each cluster ("cluster fractions"). Cluster fractions represent a patient-specific summary of histological patterns observed in the slide images.

To give an example, in a dataset of tissue biopsy slide images and associated clinical data, we can derive patient-level cluster fractions by analyzing all tile images extracted from each biopsy slide. If a specific cluster predominantly contains tile images with high-grade cancer, the fraction of this cluster for a given patient would represent the proportion of high-grade cancer observed in that patient’s biopsy slides.

\subsection{Datasets}
\label{sec:datasets}

All datasets leveraged in this study are summarized in \cref{tab:datasets}. All slide images are processed with \texttt{HistoPrep} \cite{histoprep}.

\subsubsection{Training dataset}

Training data, designated as HelsinkiProstate, for the prostate tissue encoders, consists of patients who have undergone prostate biopsies or radical prostatectomy in Helsinki University Hospital between 2013 and 2021. In total, there are 1,307 patients with 5,642 and 5,584 tissue slides of needle biopsies, and tissue sections from radical prostatectomy (\cref{tab:datasets}). To create the HelsinkiProstate dataset, each slide is cut into $640 \times 640$ pixel tile images at magnifications 20x, 10x, and 5x with 20\% overlap between neighbouring tile images. This produces several hundred million tile images, which are then preprocessed with \texttt{HistoPrep} \cite{histoprep} to remove tiles containing non-tissue areas such as pen markings or other artefacts. Once irrelevant tile images have been filtered out, all remaining tile images are run through a prostate cancer classifier model from \cite{strongaugment}. To create a balanced training dataset, all 16 million tile images with a prediction score $\hat{y}>0.2$, and 32 million randomly sampled tile images with score $\hat{y}\leq0.2$ were selected to comprise the HelsinkiProstate dataset.

\subsubsection{Evaluation datasets}

HelsinkiTMA dataset contains 1,769 TMA spots from 432 prostate cancer patients who underwent radical prostatectomy between 1983 and 1998 at the Helsinki University Hospital. All slide images are cut into $512 \times 512$ with 25\% overlap between neighbouring tiles. The patients have a median follow-up time of 19.0 years. All 432 patients have Gleason grade information available, and 238 have complete clinical data for calculating CAPRA-S \cite{capra_nomogram} score and Memorial Sloan Kettering Cancer Center 15-year survival probability \cite{mskcc_nomogram}, denoted as MSKCC-S. Kaplan-Meier survival curves \cite{kaplan_meier} for Gleason grade groups and both clinical nomograms are presented in Supplementary Fig. 7. 

Helsinki30 and Helsinki60 datasets \cite{strongaugment} contain whole slide images (normal size and whole mounts) from 30 and 60 patients who have undergone radical prostatectomy at the Helsinki University Hospital between the years 2014 and 2021. All slide images in both datasets have been annotated by pathologists, and classified as cancerous or benign. These datasets are also part of the training (HelsinkiProstate) dataset (\cref{tab:datasets}).

Several publicly available datasets are used in this study. The test set of the PESO dataset \cite{peso_dataset, peso_study} contains 137 tile images of $2,500 \times 2,500$ pixels from 37 different slide images annotated as either cancerous or benign. PANDA development dataset \cite{panda} contains 10,616 prostate biopsy slides from 2,113 patients from Radboud University Medical Center and Karolinska Institute, which we denote in this study as Radboud and Karolinska.

\subsection{Fine-tuning the encoder models}
\label{sec:fine_tuning_detauls}

To fine-tune the encoders for downstream tasks, the partial fine-tuning setup from \cite{mae} is used, with the following modifications. RandAugment \cite{randaugment} is replaced with StrongAugment \cite{strongaugment} with 2, 3 or 4 operations with 0.5, 0.3 and 0.2 probability, respectively. Each image is flipped vertically and/or horizontally and a perspective operation is applied with a scale from 0 to 0.1 with a probability of 0.5. For partial fine-tuning \cite{mae}, only the last 0.5, 2, 4 or 8 XCiT-S12 model blocks are fine-tuned, where 0.5 denotes fine-tuning the last fully connected layer of the last model block. There are 1.182, 3.550, 7.120 and 14.260 million fine-tuned parameters for 0.5, 2, 4, and 8 model blocks, and 25.868 million for the full XCiT-S12 model. For fine-tuning the encoders without training, a K-nearest neighbour (KNN) classifier is used with $k=20$. The KNN-classifier is fit with the features extracted from the training dataset, and label predictions correspond to the proportion of nearest neighbours with a positive label. Each experiment is repeated five times, and the mean and standard deviation of the area under the receiver operating curve (AUROC) are reported.

For the PESO dataset, randomly resized crops with scale [0.01, 0.2] are randomly sampled from the 2,500x2,500 pixel region of interest images during fine-tuning, and an epoch is defined as 51,200 training samples. For Karolinska and Radboud datasets, training images are randomly sampled from 384x384 pixel tile images with 20\% overlaps, and an epoch is defined as 262,144 training samples, or the maximum amount of tile images included in model training. When limiting the number of training images from the PESO dataset, region of interest images from 1, 2, 4, 8, 16 or 32 histological slides are used for training. When limiting the number of training data in the Karolinska and Radboud datasets, only 4,096, 8,192, 16,384, 32,768, 65,536, 131,072, 262,144 or 524,288 randomly selected tile images, or all 972,800 and 782,336 images in the Karolinska and Radboud datasets, are used for training. In both training data limitation experiments, the last two encoder blocks are fine-tuned.

For augmentations and transformations, \texttt{strong-augment} (0.1.0) and \texttt{albumentations} (1.3.0) \cite{albumentations} Python packages were used.

\subsubsection{Prostate cancer-specific death survival models}
\label{sec:survival_models}

To create survival models using HistoEncoder features, we first cluster the HelsinkiTMA dataset and obtain histological cluster memberships for each patient as described in \cref{sec:multimodality} with the \texttt{prostate-m} encoder and number of clusters set to 32.

Penalized Cox proportional hazards models are then used to predict prostate cancer-specific death, with a penalizer of 0.001 and an $L1$-ratio of 0.5. From the 32 patient-level clusters, six clusters are selected based on a parameter importance analysis. All models are then trained on the patient Gleason grade, CAPRA-S or MSKCC-S, with or without HistoEncoder cluster features. Each model is trained 1,000 times, where 25\% of the samples are set aside as a test set using stratified random splits. Model performance is evaluated with a concordance score, time-dependent AUC score between 1 and 23 years, and a net benefit \cite{net_benefit} curve for predicting prostate cancer-specific death at 15 years.

%% file: tables/datasets.tex
\begin{table*}
    \centering
    \caption{Datasets used in this study.}
    \small
    \begin{tabular}{lllllll}
    \toprule
        Name & Medical centre & Patients & Slides & Type & Tile images$^{*}$  & Reference \\
        \midrule
        HelsinkiProstate & Helsinki University Hospital & 1,307 & 11,226  & Biopsy \& Organ section & 898.4 million & – \\
        \midrule
        Helsinki30$^\dagger$ & Helsinki University Hospital & 30 & 465 & Organ section & 75.2 million & \cite{strongaugment} \\
        Helsinki60$^\dagger$ & Helsinki University Hospital & 60 & 863 & Organ section & 146.0 million & \cite{strongaugment} \\
        HelsinkiTMA & Helsinki University Hospital & 432 & 1,769 & Tissue microarray & 0.5 million & – \\
        PESO & Radboud University Medical Center & 40 & 137 & Region of interest & 13,287 & \cite{peso_dataset, peso_study} \\        
        Karolinska  & Karolinska Institutet & \multirow{2}{*}{2113} & 5,456 & Biopsy & 3.2 million & \cite{panda} \\
        Radboud & Radboud University Medical Center & & 5,160 & Biopsy & 2.6 million & \cite{panda} \\
    \bottomrule
    \multicolumn{6}{l}{\scriptsize{$^{*}$Non-overlapping $256\times256$ tiles with >50\% tissue; $^\dagger$Part of HelsinkiProstate dataset}}
    \end{tabular}
    \label{tab:datasets}
\end{table*}

%% file: sections/3-results.tex
\section{Results}
\label{sec:results}

\begin{figure}
    \centering
    \begin{subfigure}{\columnwidth}
        \centering
        \includegraphics[width=\columnwidth]{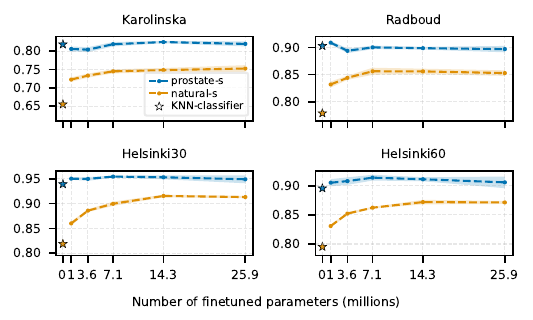}
        \subcaption{Limiting the number of fine-tuned encoder blocks.}
        \label{fig:peso_params}
    \end{subfigure}
    \hfill
    \begin{subfigure}{\columnwidth}
        \centering
        \includegraphics[width=\columnwidth]{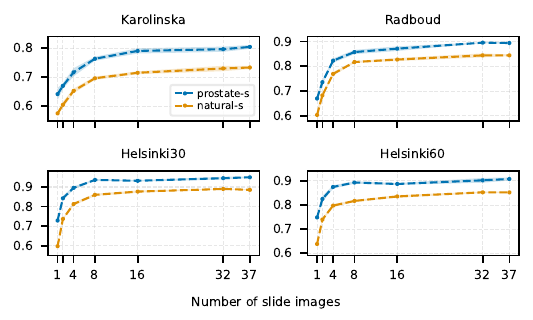}
        \subcaption{Limiting the amount of data used to train the models.}
        \label{fig:peso_images}
    \end{subfigure}
    \caption{AUROC scores for the prostate cancer classifiers fine-tuned from pre-trained \texttt{prostate-s} and \texttt{natural-s} encoder models using the PESO dataset. \texttt{Prostate-s} based models achieve higher AUROC scores than \texttt{natural-s} based models with less fine-tuned parameters (\textit{a}) and training data (\textit{b}). A simple KNN classifier requiring no training significantly outperforms a fully fine-tuned \texttt{natural-s} encoder model on all evaluation datasets.}
    \label{fig:finetune_peso}
\end{figure}

\begin{figure}
  \centering
    \includegraphics[width=\columnwidth]{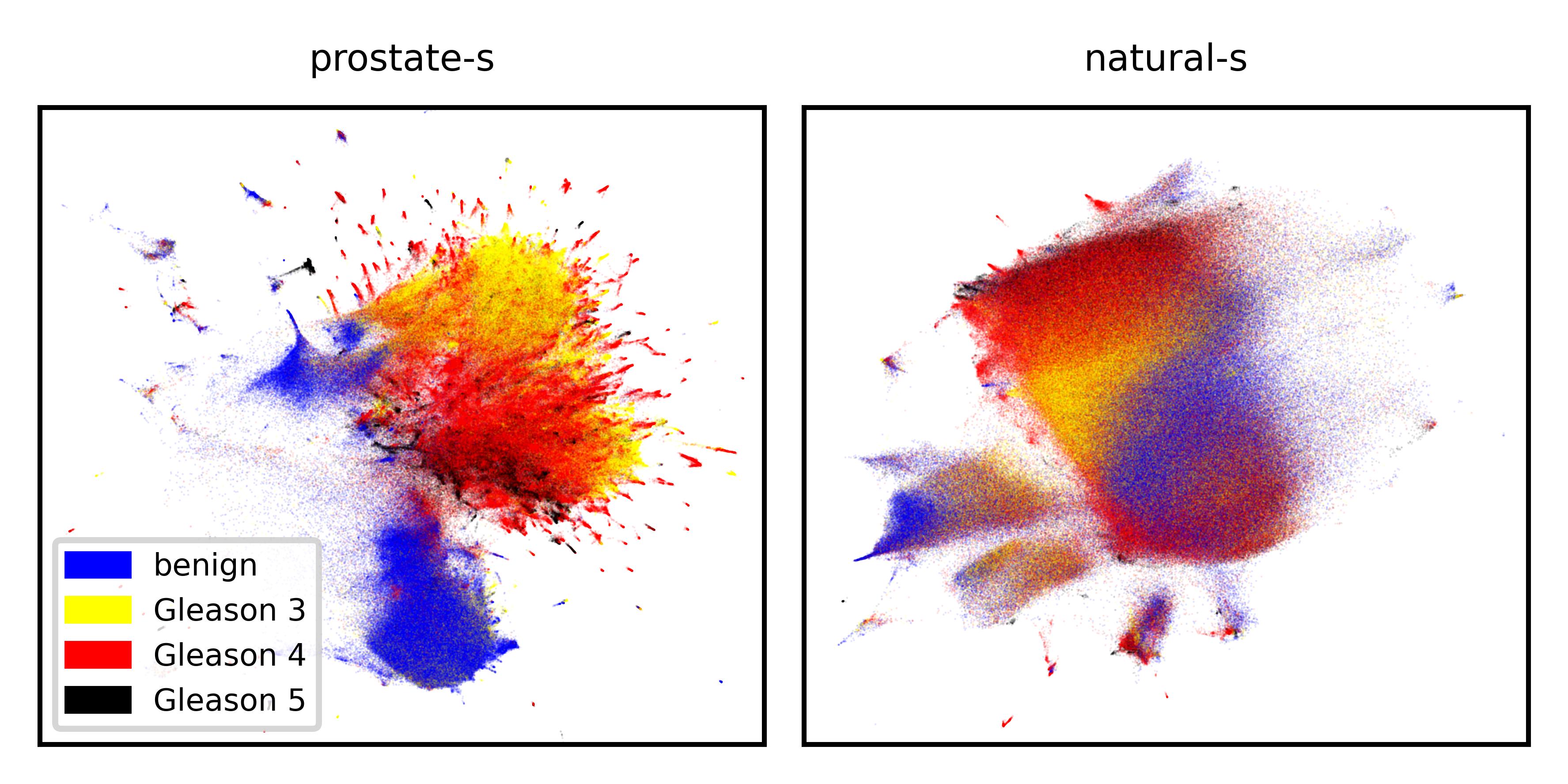}
    \caption{UMAP representation of the features extracted from the epithelium tissue in the Radboud dataset. Benign and cancerous epithelium, as well as different Gleason grades, form clear clusters in the features extracted with the \texttt{prostate-s} encoder, while features extracted with the \texttt{natural-s} model are more mixed.}
    \label{fig:umap_radboud}
\end{figure}

\begin{figure*}
    \centering
    \begin{subfigure}{\columnwidth}
        \centering
        \includegraphics[width=\columnwidth]{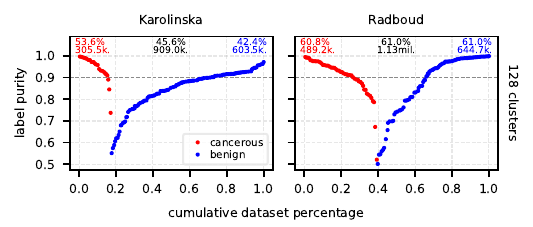}
        \subcaption{Cluster label purities the Karolinska and Radboud datasets.}
        \label{fig:cluster_purity_panda}
    \end{subfigure}
    \hfill
    \begin{subfigure}{\columnwidth}
        \centering
        \includegraphics[width=\columnwidth]{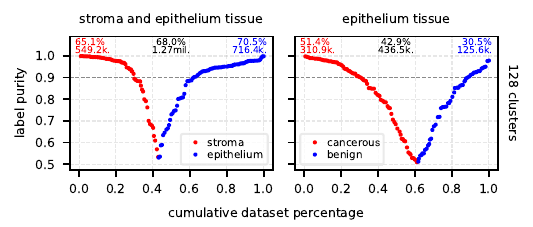}
        \subcaption{More detailed label purities for the Radboud dataset.}
        \label{fig:cluster_purity_radboud}
    \end{subfigure}
    \caption{Cluster label purities for different labels in the Karolinska and Radboud datasets. A significant proportion of tile images in these datasets are contained in clusters with greater than 90\% label purity for cancerous vs benign tissue (a), stroma vs. epithelium tissue (b left), and cancerous vs. benign epithelium tissue (b right).}
    \label{fig:cluster_purity}
\end{figure*}

\subsection{Fine-tuning a prostate cancer classifier}
\label{sec:finetune}

In this section, we report the performance of both \texttt{prostate-s} and \texttt{natural-s} encoder models in classifying prostate cancer in tissue images. The models are fine-tuned using either the PESO, or Karolinska and Radboud datasets. Both encoder models are fine-tuned while limiting either the number of fine-tuned parameters or the amount of training data. 

While limiting the number of fine-tuned parameters (\cref{fig:finetune_peso}a and Supplementary Fig. 6a)
, \texttt{prostate-s} based models significantly outperform \texttt{natural-s} based models in all evaluation datasets, regardless how many parameters are fine-tuned. \texttt{Prostate-s} based models also saturate quickly and only require minimal fine-tuning to achieve the best performance. \texttt{Prostate-s} based models seem to overfit to the training datasets (Supplementary Fig. 6a)
, and would likely benefit from higher regularisation. 

While limiting the number of training images (\cref{fig:finetune_peso}b and Supplementary Fig. 6b) 
, \texttt{prostate-s} based models significantly outperform \texttt{natural-s} based models in all evaluation datasets. \texttt{Prostate-s} based models achieve comparable or better performance with only 16 regions of interest crops from four distinct histological slides, when compared to \texttt{natural-s} based models trained on the full PESO dataset (\cref{fig:finetune_peso}b). In particular, \texttt{prostate-s} based models achieve significantly better performance with only 1,024 tile images when compared to \texttt{natural-s} based models trained with all 972,800 and 782,336 tile images in the Karolinska and Radboud datasets (Supplementary Fig. 6b),
respectively. \texttt{Prostate-s} based models also saturate quickly with diminishing returns after 8,192 tile images.

Remarkably, a KNN-classifier fitted with features extracted from \texttt{prostate-s} outperforms a fully fine-tuned \texttt{natural-s} model on all evaluation datasets (\cref{fig:finetune_peso}a and Supplementary Fig. 6).
\texttt{Prostate-s} outperform models trained in a supervised manner with ImageNet data (Supplementary Tab. 2,3, and 4).

\subsection{Identifying histological patterns as HistoEncoder feature clusters}
\label{sec:cluster_purity}

Tissue image features derived from the \texttt{prostate-s} model create clusters that clearly separate benign from cancerous epithelium (\cref{fig:umap_radboud}). This representation also allows differentiating cancer grades (Gleason). Conversely, features extracted with the natural-s model show a more diffuse pattern, making it harder to distinguish tissue types.

To quantify whether similar features are extracted for tile images with similar histological patterns, the \texttt{prostate-m} encoder is used to automatically label the Karolinska and Radboud datasets with the annotation workflow (\cref{sec:automatic_annotation}). After assigning each tile image to a cluster, the proportion of different labels in each feature cluster can be assessed. In the Karolinska dataset, 45.6\% of all tile images, 53.3\% of cancerous epithelium, and 42.4\% of benign epithelium and stroma are contained in clusters with a label purity above 90\% (\cref{fig:cluster_purity}). In the Radboud dataset, a clear majority of images belong to high-purity clusters (68.8\% all tiles, 74.7\% cancerous epithelium, and 63.9\% benign epithelium and stroma). In total, 2.1 million $384 \times 384$ tile images in the PANDA dataset could be classified into cancerous and benign tissue with over 90\% accuracy after visually inspecting only 256 tile image clusters. 

Annotations in the Radboud dataset are more granular than those in the Karolinska dataset, 
distinguishing between stroma and benign epithelium.
In the Radboud dataset, 68.0\% of all tile images, 65.1\% of stroma, and 70.5\% of epithelium are contained in clusters with a label purity above 90\% (\cref{fig:cluster_purity}b). For the harder task of classifying epithelium tissue as cancerous or benign, 42.9\% of all tile images, 51.4\% of cancerous, and 30.5\% of benign epithelium are contained in clusters with a label purity above 90\%. Varying the number of annotated clusters, we observe that as few as eight clusters are enough to find clusters with label purity over 90\% (Supplementary Fig. 8).

\subsection{Predicing prostate cancer-specific mortality}

\begin{figure*}
    \centering
    \begin{subfigure}[t]{\textwidth}
        \centering
        \includegraphics[width=\textwidth]{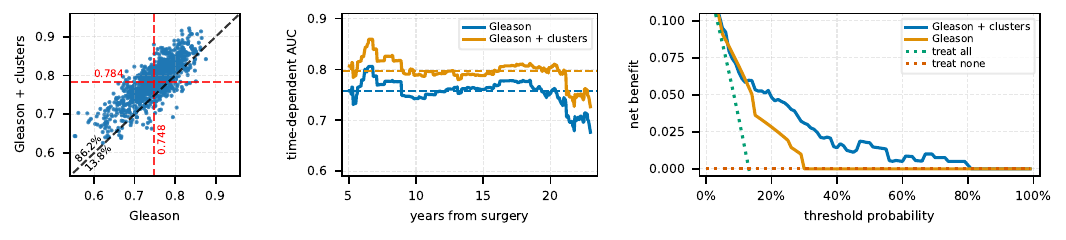}
        \subcaption{Gleason grade.}
        \label{fig:pca_death_gleason}
    \end{subfigure}
    \hfill
    \begin{subfigure}[t]{\textwidth}
        \centering
        \includegraphics[width=\textwidth]{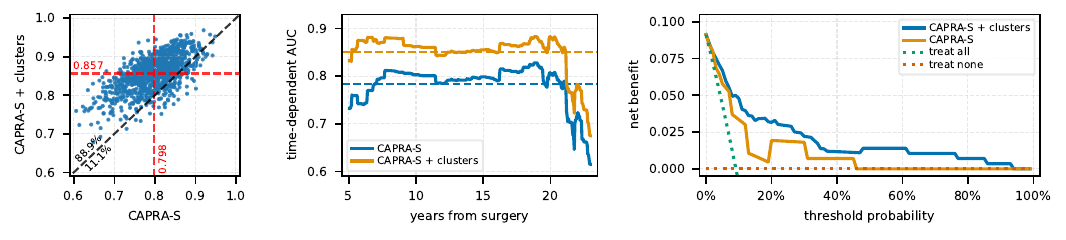}
        \subcaption{CAPRA-S risk score.}
        \label{fig:pca_death_capra}
    \end{subfigure}
    \hfill
    \begin{subfigure}[t]{\textwidth}
        \centering
        \includegraphics[width=\textwidth]{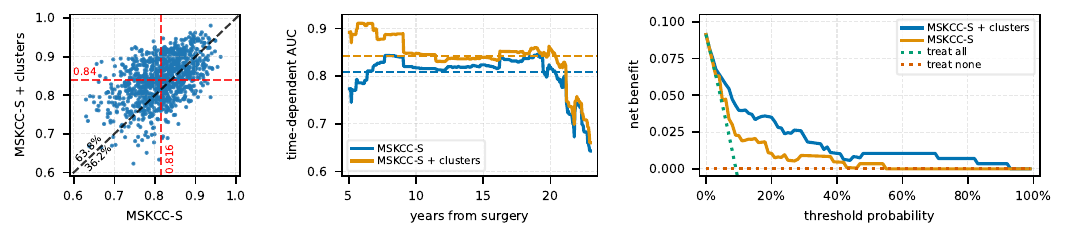}
        \subcaption{MSKCC-S prediction of being alive at 15 years.}
        \label{fig:pca_death_mskcc}
    \end{subfigure}
    \caption{Head-to-head concordance comparisons between 1,000 stratified random splits (\textit{left}), time-dependent AUC scores (\textit{centre}) and net benefit curves for prediction of prostate cancer-specific death at 15 years (\textit{right}). Including patient-level feature cluster percentages improves concordance, time-dependent AUC score and provides a higher net benefit over a wide threshold probability for the Gleason grade (\textit{a}), CAPRA-S (\textit{b}) and MSKCC-S (\textit{c}) nomograms.}
    \label{fig:pca_death}
\end{figure*}

We next combine information extracted with HistoEncoder from histological slide images with clinical data to build multimodal survival models predicting prostate cancer-specific death. The \texttt{prostate-m} encoder is used to collect patient-level feature cluster frequencies ($n=32$ clusters; \cref{sec:multimodality}) for the tissue microarray spots in the HelsinkiTMA dataset. These cluster frequencies represent the histological pattern distribution for each patient and can be simply concatenated with clinical data. Next, six feature clusters are selected and Cox proportional hazards models are fitted (see \cref{sec:survival_models}).
A visual assessment of the top six feature clusters suggests more atypia in the cellular morphologies in clusters associated with worse prognoses (Supplementary Fig. 9).
Certain subtypes and growth patterns, such as cribriform gland architecture and mucinous pattern are recurrent in the clusters with the highest HRs. Extracellular matrix formation, individual cell clusters and lymphocyte composition seem to vary even between the top six clusters.

In 1,000 random stratified splits, survival models augmented with the patient-level cluster frequencies achieve higher concordance scores than the baseline models in 84.9\%, 89.2\%, and 67.4\% of the splits with Gleason grade, CAPRA-S and MSKCC-S, respectively (\cref{fig:pca_death}, left). 
HistoEncoder-augmented models also achieve higher mean time-dependent AUC scores over a 1 to 23-year period (\cref{fig:pca_death}, center).
Finally, the augmented models yield a higher net benefit predicting prostate cancer death at 15 years over a wide threshold probability (\cref{fig:pca_death}, right).

%% file: sections/4-discussion.tex
\section{Discussion}
\label{sec:discussion}

In this work, we introduce HistoEncoder, a foundation model for prostate cancer digital histopathology. Foundation models hold significant promise to contribute to precision cancer medicine. While pre-training a foundation model takes considerable computational resources, the model can then be fine-tuned to a multitude of specific tasks~\cite{Wang2024, Truhn2024, Wojcik2022}. Importantly, fine-tuning often requires orders of magnitude less data and resources than pre-training~\cite{Wang2024,khan2024,kim2022}.

We pre-trained HistoEncoder with 48 million tile images extracted from prostate tissue images and showed how the model could distinguish malignant and benign tissues, Gleason grades, as well as stromal and epithelial tissues without any annotated information on tissue types being available during training. Previously, foundation models for digital histopathology have been typically pre-trained with natural images~\cite{he2016,szegedy2015,liu2021}. In this study, we demonstrated how pre-training with domain-specific images yielded substantially better performance than using natural images for prostate histopathology tasks. In line with our findings, recent foundation models trained on tissue images have had excellent performance in digital histopathology downstream tasks such as cancer detection and survival prediction~\cite{vorontsov2024foundation,wang2024pathology}. These models have however not challenged existing cancer-specific clinical prediction models to the extent that we address prostate cancer.

Here, we demonstrated extending and applying HistoEncoder in two clinically relevant tasks. First, we created a classifier which achieved high performance in predicting the presence of prostate cancer. We observed significantly better prediction accuracy, and compute and data efficiency with HistoEncoder models trained with tissue images compared to identical models trained with natural images, also in unsupervised cancer grading task. Second, we applied HistoEncoder in the integration of tissue imaging and clinical data to compare with commonly used nomograms for prostate cancer-specific mortality prediction. This approach resulted in a model which is able to consistently and robustly outperform the commonly applied risk classification systems in clinical use ({\em i.e.}, Gleason grading, CAPRA-S and MSKCC-S systems). Notably, only a handful of annotated cases and a personal computer were sufficient to fine-tune the HistoEncoder models for this purpose.

Previously, Pinckaers {\em et al.} developed a CNN model trained on biochemical recurrence (BCR) outcomes in a case-control setting~\cite{pinckaers2022predicting}. A biomarker derived from this ImageNet-pretrained ResNet50-D model demonstrated significant predictivity in a multivariable model including preoperative PSA, ISUP Gleason grade, pathological stage and surgical margin status. In an external validation cohort, however, the conventional ISUP Gleason grade outperformed the biomarker. This highlights the challenge of developing generalizable models for real-world clinical applications. Despite extensive data augmentation, models pre-trained with natural images may struggle to learn features that would consistently surpass Gleason's grading across diverse patient cohorts.

In contrast, our results suggest that HistoEncoder learns visualisable and comprehensible features 
(Supplementary Fig. 9) beyond Gleason-grade patterns that are useful in clinical tasks such as predicting prostate cancer mortality. There is a plethora of previous work showing that explainable features beyond conventional histopathological classification systems are prognostic or predictive of cancer-patient outcome \cite{echle2021deep,chandramouli2020computer,unger2024systematic}. However, reports visualizing comprehensible AI-derived features for a human expert and clinically meaningful multimodal approaches are scarce. Here we showed that HistoEncoder was able to learn such image features without expert guidance or labeled data. 

One of the weaknesses of our study is the lack of an external validation cohort in the survival analysis, and therefore it remains to be validated whether these features would be present also in other cohorts. Future efforts should hence evaluate whether HistoEncoder is able to extract coherent, predictive features across multiple cohorts, as this could yield a clinically applicable approach to stratify patients to subgroups based on survival probability beyond Gleason grades. 
Ultimately, complex machine learning models need to provide additional clinical value beyond current classification systems such as Gleason scoring, preferably with explainable features, in order to be adopted in everyday clinical practice.

\subsection{Extending HistoEncoder}
We provide HistoEncoder as an easy-to-use Python package containing both the models pre-trained with prostate tissue images, as well as a standalone software tool to extract features and cluster tissue tile images. HistoEncoder can thus be extended to specific tasks involving any histopathological images. 

As proof of concept, we were able to address cancer detection and mortality prediction with HistoEncoder using limited data and computing resources from a single laptop computer. This exemplifies the utility of such foundation models in lowering the barrier to creating clinical software tools. The \texttt{prostate-s} and \texttt{prostate-m} encoder models were able to compress information from large tissue areas into a feature vector, which was subsequently combined with other data modalities. If the other data modalities, such as spatial transcriptomics data~\cite{staahl2016} or TMA spot labels, contain information for small tissue regions, feature vectors can be directly extracted for these regions with HistoEncoder models. 

Taken together, our results demonstrate how HistoEncoder can provide clinically relevant insights from tissue images. We also highlight the importance of training data representing domain-specific data. Foundation models such as HistoEncoder allow computational methods to be quickly developed for precision cancer medicine tasks with small amounts of domain-specific data. In this study, we did not evaluate whether HistoEncoder would be useful as a foundation for models involving other cancers than prostate cancer, or in a pan-cancer setting. An exciting direction would be to explore the utility of foundation models in the analysis of multiple clinically relevant modalities and as part of multimodal predictor and interpreter models~\cite{zhou2023,song2023,moor2023}. It is crucial to evaluate HistoEncoder against tissue-agnostic foundation models pre-trained specifically on histopathology images~\cite{chen2024towards, wang2024pathology, vorontsov2024foundation}. Future multimodal models are likely to allow leveraging tissue imaging on a large scale in conjunction with other clinical and molecular profiling data, leading to improved tools for cancer diagnosis, prognosis and treatment.

%% file: extras.tex
\section*{Acknowledgements}
This work was supported by Cancer Foundation Finland [304667, 191118], Jane and Aatos Erkko Foundation [290520], Research Council of Finland [322675] and Hospital District of Helsinki and Uusimaa [TYH2018214, TYH2018222, TYH2019235, TYH2019249]. The authors also wish to acknowledge FIMM Digital Microscopy and Molecular Pathology Unit supported by HiLIFE and Biocenter Finland for imaging services, and CSC – IT Center for Science, Finland for generous computational resources on the LUMI supercomputer (LUMI Extreme Scale project).

\section*{Ethics statement}
Ethical approvals for the use of human tissue material and clinicopathologic data were obtained from the Institutional Ethics Committee of Hospital District of Helsinki and Uusimaa (§70/16.5.2018; HUS/419/2018) and by the National Supervisory Authority for Welfare and Health (VALVIRA, D:no V/38176/2018). According to the national and European Union legislation on noninterventional medical research, the study was conducted without informed individual patient consent by permission of the Hospital District of Helsinki and Uusimaa (§105/21.12.2018; HUS/419/2018). The experiments conformed to the principles set out in the WMA Declaration of Helsinki and the Department of Health and Human Services Belmont Report.

\section*{Data availability}
The original human subject and sample-related data is available upon request, provided the institutional ethical approval and research permit allow data sharing according to the particular request. Restrictions may apply to the availability of the internal Helsinki University Hospital datasets, which cannot be made publicly available due to general data protection regulations, national legislation and institutional guidelines. Publicly available datasets can be accessed through their respective publications. For data inquiries, please contact \texttt{tuomas.mirtti@helsinki.fi}.

\section*{Declaration of competing interests}
The authors have no interests to declare.

%% file: supplementary.tex
\clearpage
\setcounter{page}{1}

\input{tables/knn_tables}

\input{tables/cox_models}
\clearpage

\begin{figure*}
    \centering
    \begin{subfigure}[t]{0.7\columnwidth}
        \centering
        \includegraphics[width=\columnwidth]{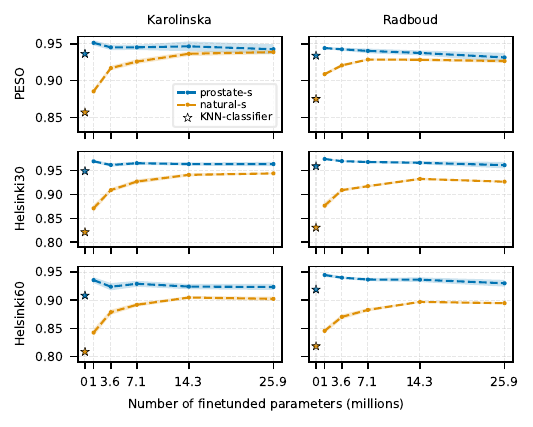}
        \subcaption{Limiting the number of fine-tuned encoder blocks.}
        \label{fig:panda_params}
    \end{subfigure}
    \hfill
    \begin{subfigure}[t]{0.7\columnwidth}
        \centering
        \includegraphics[width=\columnwidth]{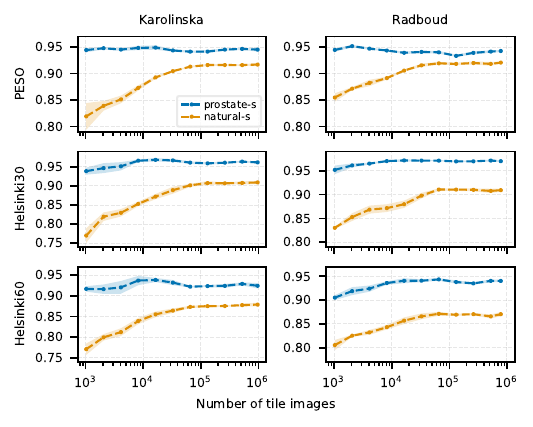}
        \subcaption{Limiting the amount of training data.}
        \label{fig:panda_images}
    \end{subfigure}
    \caption{AUROC scores for the prostate cancer classifiers fine-tuned from \texttt{prostate-s} and \texttt{natural-s} encoder models using the Karolinska and Radboud datasets. \texttt{Prostate-s} based models achieve higher AUROC scores than \texttt{natural-s} based models with less fine-tuned parameters (\textit{a}) and training data (\textit{b}). A KNN-classifier achieves similar or significantly better performance than fully fine-tuned \texttt{natural-s} based models. When limiting the number of tile images (\textit{b}), \texttt{prostate-s} based models significantly outperform \texttt{natural-s} based models even trained with 1000 times less data.}
    \label{fig:finetune_panda}
\end{figure*}

\begin{figure*}[!h]
    \centering
    \includegraphics{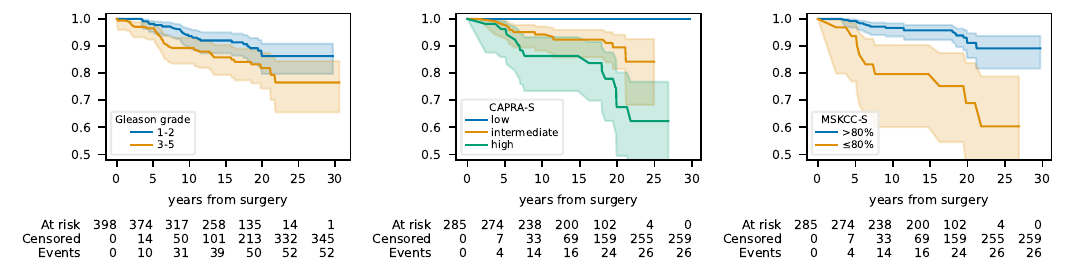}
    \caption{Kaplan-Meier survival curve estimations for Gleason grades and clinical nomograms in HelsinkiTMA cohort used for the prostate cancer death
survival models.}
    \label{fig:kaplan_meier}
\end{figure*}

\begin{figure*}[h]
    \centering    \includegraphics{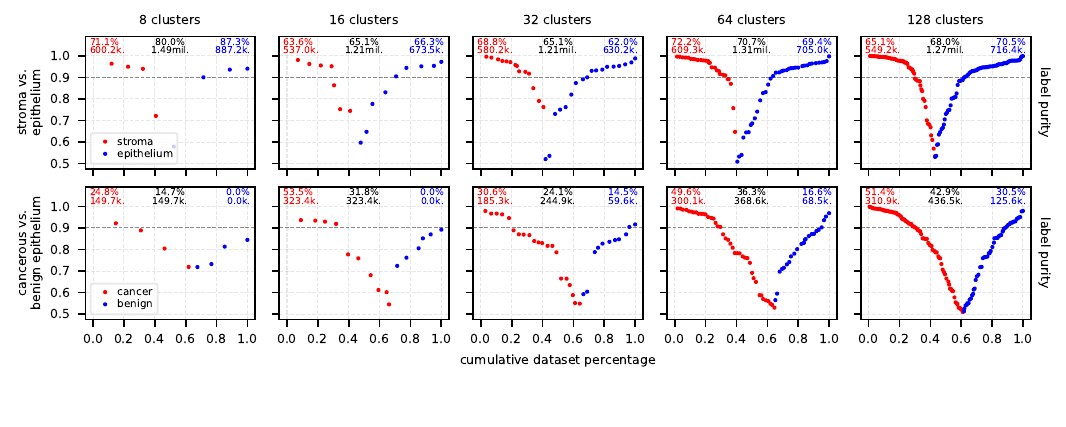}
    \caption{Cluster purities for the Radboud dataset with a varying number of clusters.}
    \label{fig:radboud_purity_multi}
\end{figure*}

\begin{figure*}[h]
    \centering
    \includegraphics{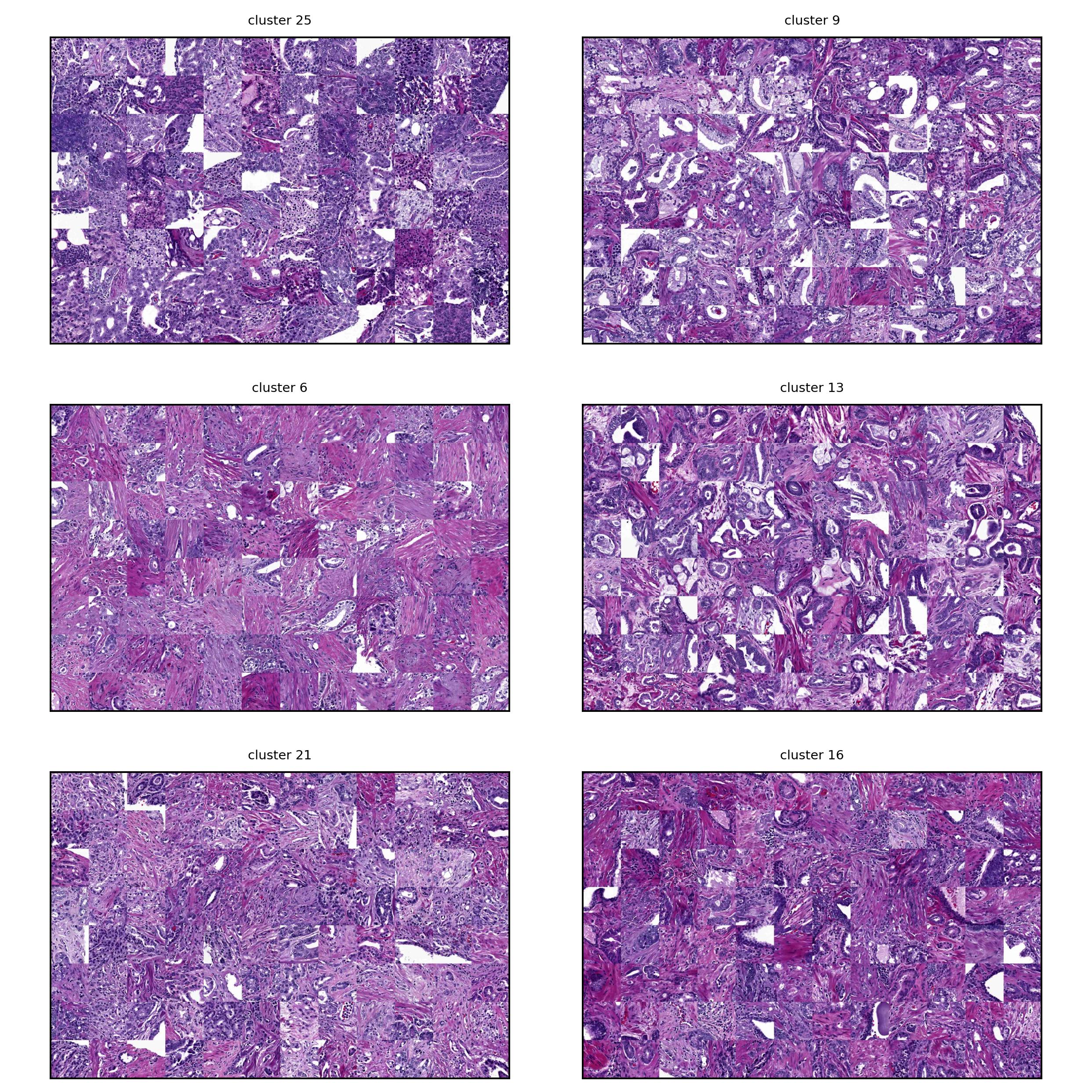}
    \caption{Randomly sampled tile images from the clusters used in the prostate cancer death survival models.}
    \label{fig:cluster_tile_images}
\end{figure*}

%% file: tables/knn_tables.tex
\begin{table*}
    \centering
    \caption{AUROC scores for a K-nearest neighbor prostate cancer classifier, fitted with features extracted from the PESO dataset.}
    \small
    \begin{tabular}{llllllll}
        \toprule
        Model & Parameters & Training method & Training data & Karolinska & Radboud & Helsinki30 & Helsinki60 \\
        \midrule
        XCiT-S12 & 25.9 & self-supervised & HelsinkiProstate & 0.819 & \textbf{0.903} & 0.939 & 0.896 \\
        XCiT-M24 & 83.9 & self-supervised & HelsinkiProstate & \textbf{0.824} & 0.898 & \textbf{0.943} & \textbf{0.901} \\
        \midrule
        XCiT-S12 & 25.9 & self-supervised & ImageNet & 0.655 & 0.779 & 0.819 & 0.795 \\
        XCiT-M24 & 83.9 & self-supervised & ImageNet & 0.683 & 0.790 & 0.816 & 0.806 \\
        \midrule
        EfficientNet-B0 & 4.0 & supervised & ImageNet & 0.668 & 0.774 & 0.550 & 0.552 \\
        EfficientNet-B1 & 6.5 & supervised & ImageNet & 0.666 & 0.753 & 0.545 & 0.54 \\
        EfficientNet-B2 & 7.7 & supervised & ImageNet & 0.671 & 0.767 & 0.558 & 0.563 \\
        EfficientNet-B3 & 10.7 & supervised & ImageNet & 0.661 & 0.740 & 0.549 & 0.548 \\
        EfficientNet-B4 & 17.5 & supervised & ImageNet & 0.652 & 0.747 & 0.547 & 0.553 \\
        ResNet-18 & 11.2 & supervised & ImageNet & 0.682 & 0.762 & 0.547 & 0.553 \\
        ResNet-50 & 23.5 & supervised & ImageNet & 0.684 & 0.746 & 0.525 & 0.521 \\
        ResNet-101 & 42.5 & supervised & ImageNet & 0.618 & 0.676 & 0.531 & 0.531 \\
        XCiT-S12 & 25.9 & supervised & ImageNet & 0.645 & 0.690 & 0.707 & 0.714 \\
        XCiT-M24 & 83.9 & supervised & ImageNet & 0.647 & 0.720 & 0.752 & 0.739 \\
        \bottomrule
    \end{tabular}
    \label{tab:knn_peso}
\end{table*}

\begin{table*}
    \centering
    \caption{AUROC scores for a K-nearest neighbor prostate cancer classifier, fitted with features extracted from the Karolinska dataset.}
    \small
    \begin{tabular}{llllllll}
        \toprule
        Model & Parameters & Training method & Training data & PESO & Radboud & Helsinki30 & Helsinki60 \\
        \midrule
        XCiT-S12 & 25.9 & self-supervised & HelsinkiProstate & 0.936 & 0.916 & \textbf{0.949} & \textbf{0.908} \\
        XCiT-M24 & 83.9 & self-supervised & HelsinkiProstate & \textbf{0.940} & \textbf{0.918} & 0.945 & 0.902\\
        \midrule
        XCiT-S12 & 25.9 & self-supervised & ImageNet & 0.857 & 0.811 & 0.821 & 0.808 \\
        XCiT-M24 & 83.9 & self-supervised & ImageNet & 0.837 & 0.790 & 0.796 & 0.790 \\
        \midrule
        EfficientNet-B0 & 4.0 & supervised & ImageNet & 0.685 & 0.778 & 0.526 & 0.544 \\
        EfficientNet-B1 & 6.5 & supervised & ImageNet & 0.707 & 0.778 & 0.529 & 0.551 \\
        EfficientNet-B2 & 7.7 & supervised & ImageNet & 0.692 & 0.770 & 0.523 & 0.544 \\
        EfficientNet-B3 & 10.7 & supervised & ImageNet & 0.744 & 0.784 & 0.541 & 0.551 \\
        EfficientNet-B4 & 17.5 & supervised & ImageNet & 0.704 & 0.749 & 0.541 & 0.547 \\
        ResNet-18 & 11.2 & supervised & ImageNet & 0.734 & 0.770 & 0.531 & 0.553 \\
        ResNet-50 & 23.5 & supervised & ImageNet & 0.710 & 0.718 & 0.529 & 0.530\\
        ResNet-101 & 42.5 & supervised & ImageNet & 0.643 & 0.706 & 0.511 & 0.522\\
        XCiT-S12 & 25.9 & supervised & ImageNet & 0.784 & 0.683 & 0.664 & 0.648 \\
        XCiT-M24 & 83.9 & supervised & ImageNet & 0.736 & 0.738 & 0.664 & 0.631 \\
        \bottomrule
    \end{tabular}
    \label{tab:knn_karo}
\end{table*}

\begin{table*}
    \centering
    \caption{AUROC scores for a K-Nearest neighbor prostate cancer classifier, fitted with features extracted from the Radboud dataset.}
    \small
    \begin{tabular}{llllllll}
        \toprule
        Model & Parameters & Training method & Training data & PESO & Karolinska & Helsinki30 & Helsinki60 \\
        \midrule
        XCiT-S12 & 25.9 & self-supervised & HelsinkiProstate & \textbf{0.934} & \textbf{0.854} & \textbf{0.959} & \textbf{0.919} \\
        XCiT-M24 & 83.9 & self-supervised & HelsinkiProstate & \textbf{0.934} & 0.852 & 0.956 & 0.910 \\
        \midrule
        XCiT-S12 & 25.9 & self-supervised & ImageNet & 0.875 & 0.757 & 0.831 & 0.818 \\
        XCiT-M24 & 83.9 & self-supervised & ImageNet & 0.865 & 0.756 & 0.819 & 0.804 \\
        \midrule
        EfficientNet-B0 & 4.0 & supervised & ImageNet & 0.768 & 0.740 & 0.545 & 0.545 \\
        EfficientNet-B1 & 6.5 & supervised & ImageNet & 0.764 & 0.736 & 0.538 & 0.537 \\
        EfficientNet-B2 & 7.7 & supervised & ImageNet & 0.745 & 0.721 & 0.537 & 0.537 \\
        EfficientNet-B3 & 10.7 & supervised & ImageNet & 0.773 & 0.718 & 0.559 & 0.550 \\
        EfficientNet-B4 & 17.5 & supervised & ImageNet & 0.676 & 0.653 & 0.499 & 0.501 \\
        ResNet-18 & 11.2 & supervised & ImageNet & 0.786 & 0.737 & 0.567 & 0.569 \\
        ResNet-50 & 23.5 & supervised & ImageNet & 0.780 & 0.711 & 0.551 & 0.543 \\
        ResNet-101 & 42.5 & supervised & ImageNet & 0.690 & 0.683 & 0.517 & 0.526 \\
        XCiT-S12 & 25.9 & supervised & ImageNet & 0.805 & 0.714 & 0.763 & 0.746 \\
        XCiT-M24 & 83.9 & supervised & ImageNet & 0.797 & 0.723 & 0.776 & 0.732 \\
        \bottomrule
    \end{tabular}
    \label{tab:knn_radb}
\end{table*}

%% file: tables/cox_models.tex
\begin{table*}[t]
    \centering
    \caption{Cox proportional hazards model coefficients and p-values with Gleason grade and HistoEncoder feature cluster frequencies.}
    \label{tab:cox_gleason}
    \small
    \begin{tabular}{llllllll}
        \toprule
        Variable & coefficient & exp(coefficient) & SE(coefficient) & CI, lower 95\% & CI, upper 95\% & $p$ \\
        \midrule
        GG1 (ref.) &  &  &  &  &  &  \\
        GG2 & 2.01 & 7.45 & 0.83 & 0.39 & 3.63 & 0.02  \\
        GG3 & 1.19 & 3.29 & 0.88 & -0.54 & 2.92 & 0.18  \\
        GG4 & 2.59 & 13.31 & 0.84 & 0.94 & 4.24 & <0.005  \\
        GG5 & 1.91 & 6.74 & 0.95 & 0.04 & 3.78 & 0.05  \\
        Cluster 25 & 3.16 & 23.47 & 0.70 & 1.79 & 4.52 & <0.005  \\
        Cluster 9 & 3.06 & 21.40 & 0.97 & 1.16 & 4.97 & <0.005  \\
        Cluster 6 & 2.64 & 13.98 & 1.39 & -0.09 & 5.36 & 0.06  \\
        Cluster 13 & -1.78 & 0.17 & 2.60 & -6.88 & 3.32 & 0.49  \\
        Cluster 21 & 2.41 & 11.09 & 1.07 & 0.32 & 4.49 & 0.02  \\
        Cluster 16 & 4.89 & 133.20 & 1.61 & 1.74 & 8.05 & <0.005  \\
        \bottomrule
        \multicolumn{6}{l}{\scriptsize{Concordance=0.81; L1-ratio=0.5; penalizer=0.001; observations=398; events=52}}
    \end{tabular}
\end{table*}

\begin{table*}[t]
    \centering
    \caption{Cox proportional hazards model coefficients and $p$-values with CAPRA-S nomogram and HistoEncoder feature cluster frequencies.}
    \label{tab:cox_capra}
    \small
    \begin{tabular}{llllllll}
        \toprule
        Variable & coefficient & exp(coefficient) & SE(coefficient) & CI, lower 95\% & CI, upper 95\% & $p$ \\
        \midrule
        CAPRA-S & 0.45 & 1.58 & 0.11 & 0.24 & 0.67 & <0.005 \\
        Cluster 25 & 0.60 & 1.82 & 0.13 & 0.34 & 0.85 & <0.005 \\
        Cluster 9 & 0.43 & 1.54 & 0.11 & 0.21 & 0.66 & <0.005 \\
        Cluster 6 & 0.34 & 1.40 & 0.12 & 0.10 & 0.58 & 0.01 \\
        Cluster 13 & 0.07 & 1.07 & 0.27 & -0.46 & 0.59 & 0.80 \\
        Cluster 21 & 0.18 & 1.20 & 0.14 & -0.09 & 0.45 & 0.19 \\
        Cluster 16 & 0.33 & 1.39 & 0.34 & -0.34 & 1.00 & 0.33 \\
        \bottomrule
        \multicolumn{5}{l}{\scriptsize{Concordance=0.89; L1-ratio=0.5; penalizer=0.001; observations=285; events=26}}
    \end{tabular}
\end{table*}

\begin{table*}[t]
    \centering
    \caption{Cox proportional hazards model coefficients and $p$-values with MSKCC-S nomogram and feature cluster frequencies.}
    \label{tab:cox_mskcc}
    \small
    \begin{tabular}{llllllll}
        \toprule
        Variable & coefficient & standard deviation & CI, lower 95\% & CI, upper 95\% & $p$\\
        \midrule
        MSKCC-S & -0.54 & 0.58 & 0.14 & -0.82 & -0.26 & <0.005 \\
        Cluster 25 & 0.57 & 1.76 & 0.12 & 0.32 & 0.81 & <0.005 \\
        Cluster 9 & 0.42 & 1.53 & 0.11 & 0.20 & 0.65 & <0.005 \\
        Cluster 6 & 0.40 & 1.48 & 0.12 & 0.16 & 0.63 & <0.005 \\
        Cluster 13 & 0.01 & 1.01 & 0.26 & -0.50 & 0.52 & 0.98 \\
        Cluster 21 & 0.17 & 1.19 & 0.13 & -0.09 & 0.43 & 0.20 \\
        Cluster 16 & 0.30 & 1.35 & 0.36 & -0.40 & 1.00 & 0.40 \\
        \bottomrule
        \multicolumn{5}{l}{\scriptsize{Concordance=0.86; L1-ratio=0.5; penalizer=0.001; observations=285; events=26}}
    \end{tabular}
\end{table*}